\documentclass{aa}  
\def\lsi{LS~I$+61^\circ303$}
\def\e{$\pm$}  
\def\kms{km~s$^{-1}$}

\def\Ha{H$\alpha$}

\usepackage{graphicx}
\usepackage{txfonts}
\begin{document}

\title{Connection between orbital modulation of  H$\alpha$ and gamma-rays in the Be/X-ray binary \lsi }
   \titlerunning{LSI+61$^0$303: connection between \Ha\ emission, optical V, and gamma-ray emission}
   \authorrunning{Zamanov et al. }
   \author{R. Zamanov\inst{1}
          \and J. Mart{\'{\i}}\inst{2}  
	  \and   K. Stoyanov\inst{1}	
	  \and   A. Borissova\inst{1} 
 	  \and   N. A. Tomov\inst{1} 
         }  
   \institute{Institute of Astronomy and National Astronomical Observatory, 
              Bulgarian Academy of Sciences, 72 Tsarigradsko Shose Blvd., 1784 Sofia, Bulgaria
         \\   \email{rkz@astro.bas.bg, jmarti@ujaen.es,  kstoyanov@astro.bas.bg,  tomov@astro.bas.bg}
	 \and
           Departamento de F\'isica (EPSJ), Universidad de Ja\'en, Campus Las Lagunillas,  A3-420, 23071, Ja\'en, Spain 
          \\
             }
   \date{Received  November 1, 2013; accepted   ......  .., 2014}
 \abstract{We studied the average orbital modulation of various parameters ($\gamma$-ray flux, 
           \Ha\ emission line, optical V band brightness)  of the radio- and  $\gamma$-ray emitting  Be/X-ray binary   \lsi.
     Using the Spearman rank correlation test, we  found highly significant correlations between the orbital variability of
     the equivalent width of the blue hump of the \Ha\ and  $Fermi$-LAT flux  with a Spearman p-value $\sim \! 2 \times 10^{-5}$, 
     and  the equivalent widths ratio $EW_B/EW_R$ and $Fermi$-LAT flux  with  p-value $\sim \! 9 \times 10^{-5}$.
     We also found a significant anti-correlation between  $Fermi$-LAT flux  and V band magnitude  
     with p-value $\sim \!7 \times  10^{-4}$. \\     
     All these correlations refer to the average orbital variability, and we conclude 
     that the \Ha\ and $\gamma$-ray  emission processes in \lsi\ are  connected.  
     The possible physical scenario is briefly discussed.  
    }
   \maketitle
%

\section{Introduction}  
\label{intro}

\lsi\ (V615~Cas)  is a high-mass X-ray binary whose remarkable nature was illustrated  following the discovery of its strong,
non-thermal, and periodic radio outbursts  (Gregory \& Taylor 1978). 
It was first detected  as a strong  $\gamma$-ray source  by the {\it COS~B} satellite (Hermsen et al. 1977). 
More recently, it has also been reported as a source of  high-energy (HE) and very high-energy (VHE)
$\gamma$-rays by the {\it Fermi} Large Area Telescope (LAT, Abdo et al. 2009),  the
MAGIC Cherenkov telescope (Albert et al. 2006), and  by the VERITAS collaboration (Maier et al. 2012). 
Indeed, it is currently considered one of the few confirmed representatives of the
selected class of $\gamma$-ray binaries since the system's luminosity in this energy range  dominates  the whole spectral energy distribution.

\lsi\ consists of  a massive  B0Ve star and a compact object orbiting the primary every 26.5 d.
According to the most recent radial velocity measurements of the 
absorption lines of the primary (Casares et al. 2005, Aragona et al. 2009), the orbit is elliptical ($e=0.537\pm0.034$),
with periastron passage determined to occur around phase $\phi=0.275$. 
The compact object interacts with the Be  circumstellar disk thereby  sampling a wide range of physical parameters and producing remarkable,  
periodic flaring events each orbital cycle. Such a strong orbital
modulation in the  \lsi\ emission is observed across the whole electromagnetic spectrum, especially in the radio (Taylor et al. 1992), 
 optical  (Mendelson \& Mazeh 1994), X-ray (Paredes et al. 1997, Leahy 2001), HE (Abdo et al. 2009), and VHE $\gamma$-ray (Albert et al. 2009) domains.
 In the optical, the orbital period signature is evident not only in visible broad band photometry, but also in the
 spectral properties of the \Ha\ emission line (Zamanov et al. 1999; Grundstrom et al. 2007).
 The scenario of compact companion interaction with the Be disk is currently favored by Very Long Baseline Interferometry (VLBI) 
 images that show a cometary structure on milli-arcsecond angular scales that rotates with the orbital period (Dhawan et al. 2006).
 
In addition to the orbital periodicity, another clock is operating in the system. 
A periodic modulation of about 4.4 yr in the phase and amplitude of the radio  outbursts was first reported by Paredes (1987) and Gregory et al. (1989).
This super-orbital modulation has also been detected in \Ha\ (Zamanov et al.  1999), 
X-rays (Li et al. 2012), and $\gamma$-rays (Ackermann et al. 2013). It could be due to precession 
of the Be disk (Lipunov \& Nazin 1994), 
a beat frequency between the orbital and precessional rates (Massi \& Jaron 2013), or  
quasi-cyclical variability of the equatorial outflow of the Be star. 

Previous multiwavelength observations have revealed interesting correlations between the X-ray and VHE gamma-ray flares (Albert et al. 2008, Anderhub et al. 2009).
This suggests that the same relativistic electrons that radiate inverse Compton VHE photons also produce synchrotron X-ray emission.
In this Letter, we further explore the multiwavelength behavior of \lsi\ in different spectral domains and search for correlations among them
that could help to better characterize the physical mechanism behind the system's orbital flaring episodes. In particular, we focus our attention
on the two \lsi\ extensive observational monitorings that are currently available, namely in $\gamma$-rays and in H$\alpha$ high-resolution spectroscopy.
Other observational databases in the radio and optical domains are also included in our study.

\section{Observations}
To create the folded light curve of \lsi,  we used  the following data:   

For {\bf HE $\gamma$-rays, } the {\it Fermi}
team monitors flux values for a number of bright sources and transient sources that 
cross their monitoring flux threshold.
Here we downloaded the {\it Fermi} LAT daily-averaged flux values for \lsi\ in the energy range from 0.1 to 300 GeV.
At the time of writing, this data set covered the time interval from JD 2454688.5 (2008 August) to JD 2456358.5 (2013 March).


{\bf  Johnson  V band } magnitudes were taken from Lipunova (1988), Paredes et al. (1994), Zaitseva \& Borisov (2003).
Among the photometry available in the literature, we used only those observations that are reduced to Johnson's system. 
We also calculated average orbital variability using the  unfiltered optical magnitudes (149 measurements)
from  the Northern Sky Variability Survey (Wo{\'z}niak et al. 2004). 
These unfiltered magnitudes are not plotted here, but they do confirm the detected V band  variability. 

{\bf  \Ha\ spectroscopic data} were taken from 
Paredes et al. (1994), Steele et al. (1996), 
Liu \& Yan (2005), Grundstrom et al. (2007), McSwain et al. (2010), 
and Zamanov et al. (1999, 2013). 
Among the  various parameters of the \Ha\ emission line we used here  are 
the total equivalent width of the H$\alpha$ emission line, hereafter $EW$, 
the equivalent width of the blue hump EW(B),
the equivalent width of the red hump EW(R),
the ratio between the equivalent widths of the blue and red humps $EW_B/EW_R$,
and the distance between the peaks, $\Delta V_p$.  

In {\bf radio photometry,} flux densities were retrieved from the old Green Bank Interferometer (GBI), which is a facility
of the USA National Science Foundation operated by NRAO in support of the NASA High Energy Astrophysics program.
A total of  7234 observations of the flux density at  2.25 GHz  and 8.3 GHz obtained  
from JD2450410 (November 1996)   untill   2451664 (April 2000) are available for study.

 \begin{figure}  
  \vspace{23.0cm}    
  \includegraphics{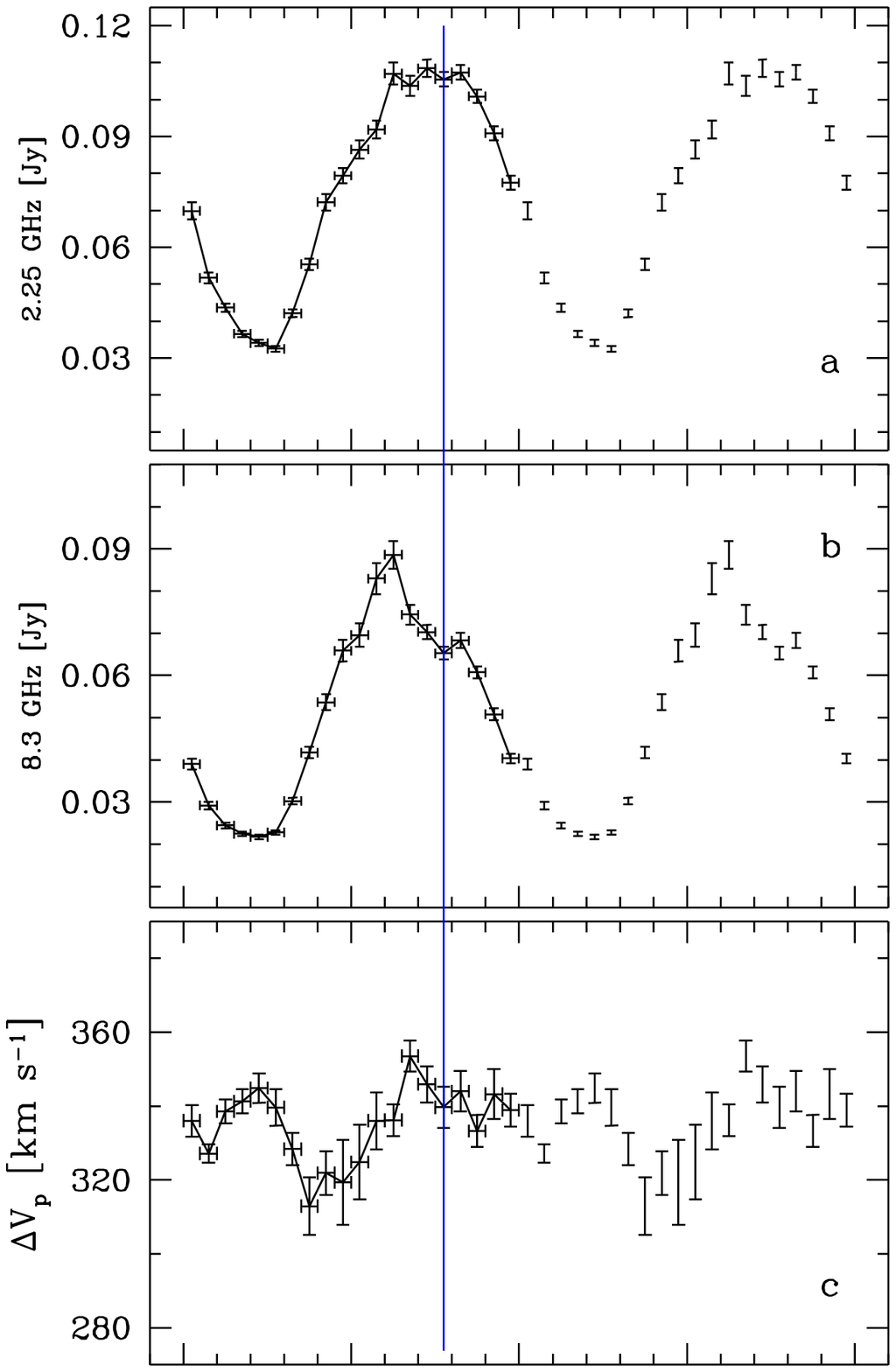}    
  \includegraphics{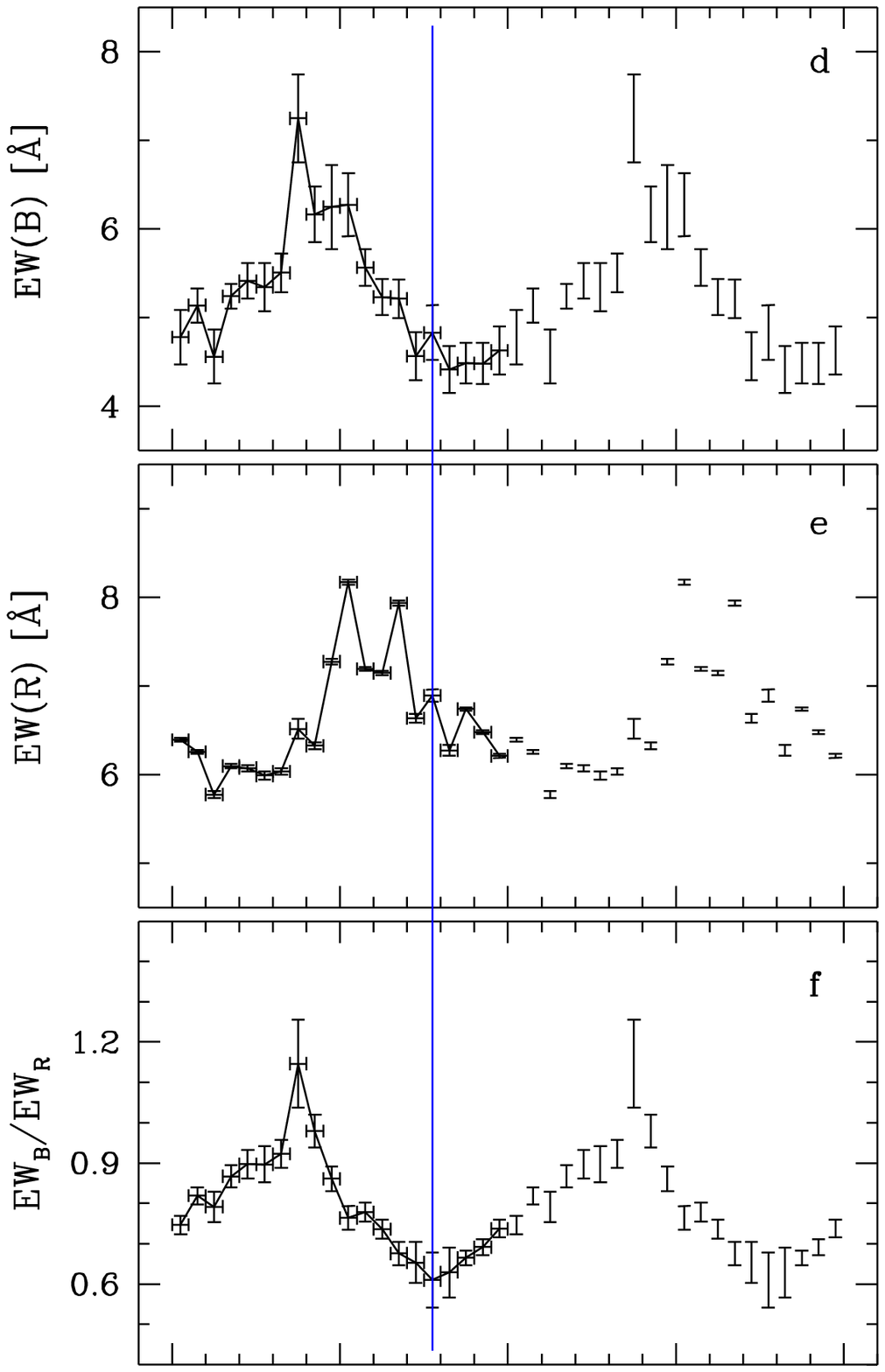}    
  \includegraphics{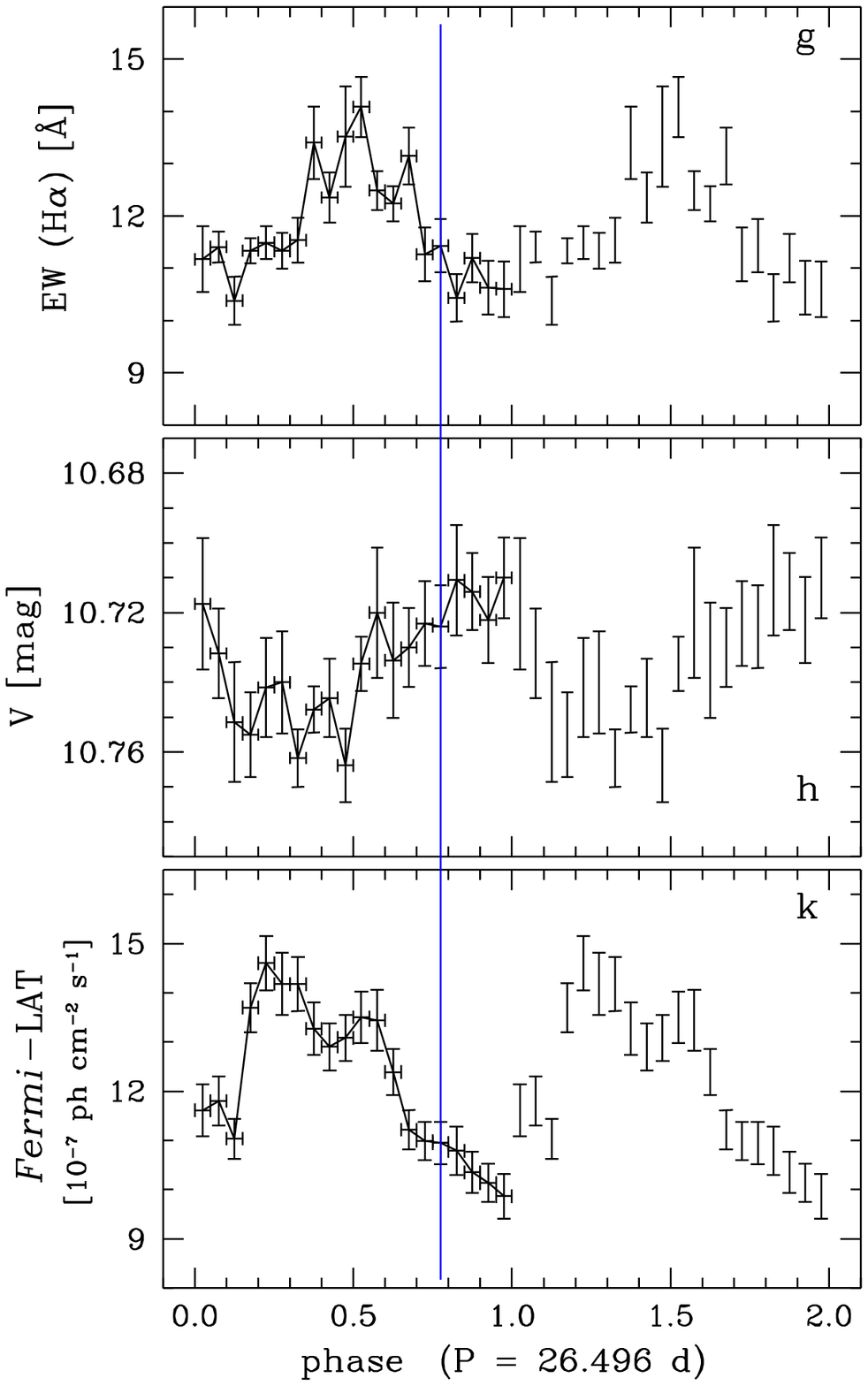}    
  \caption[]{The averaged  orbital variability of \lsi.
     Plotted from up to down: 
     the radio fluxes at 2.25 and 8.3 GHz,  
     $\Delta V_p$,
     EW(B),
     EW(R),
     $EW_B/EW_R$,
     EW(\Ha), 
     the optical V,  
     and the $Fermi$-LAT flux. 
     The vertical (blue) line indicates the apastron passage. 
     The data are  averaged in 20 bins that are 0.05 each.
  }  
\label{f1...}      
\end{figure}	     

 \begin{figure}   
  \vspace{23.3cm}   
  \includegraphics{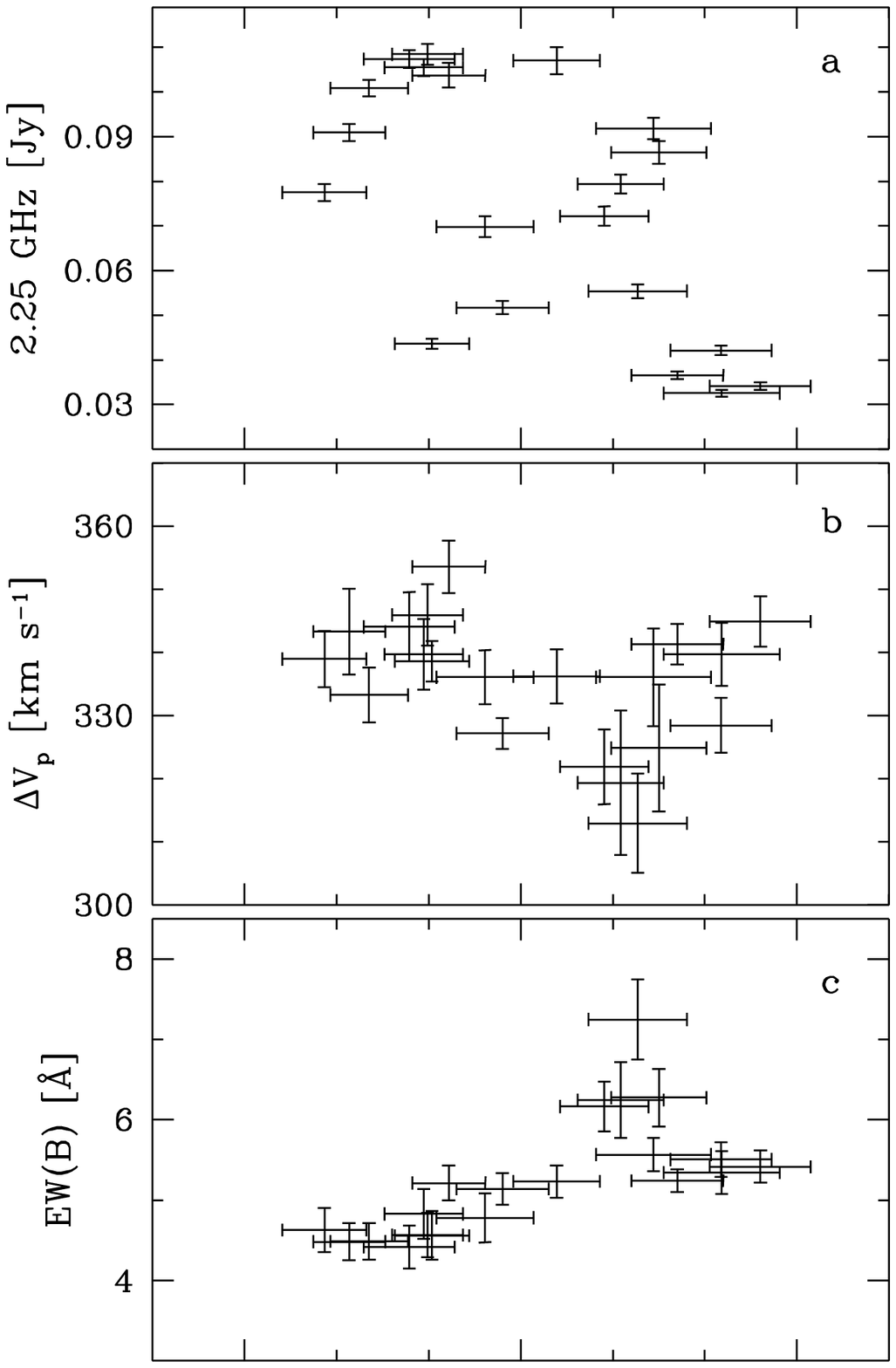}      
  \includegraphics{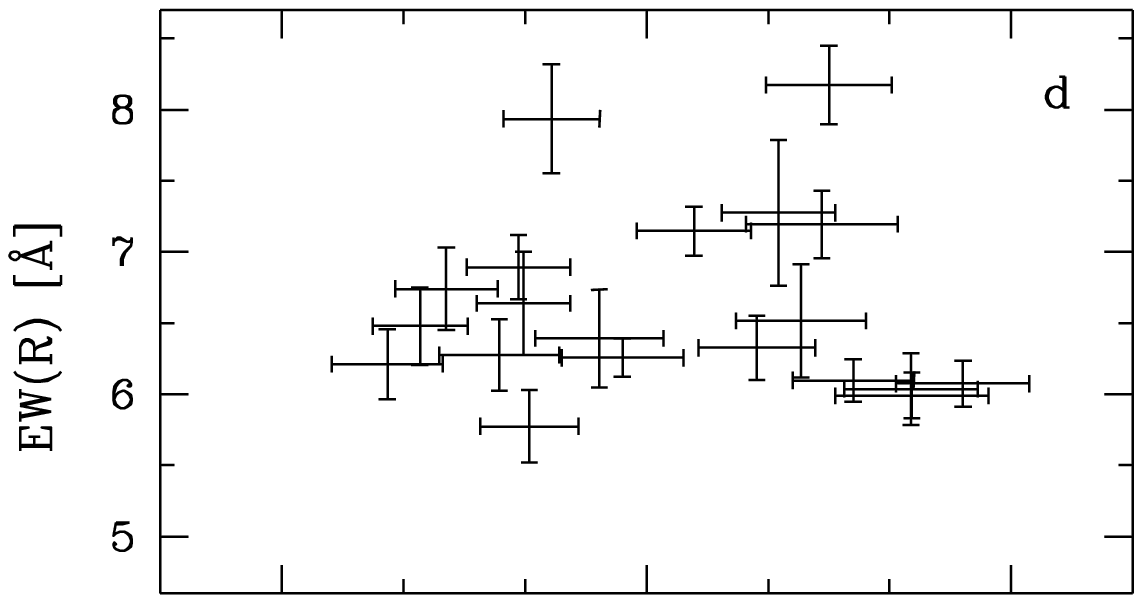}      
  \includegraphics{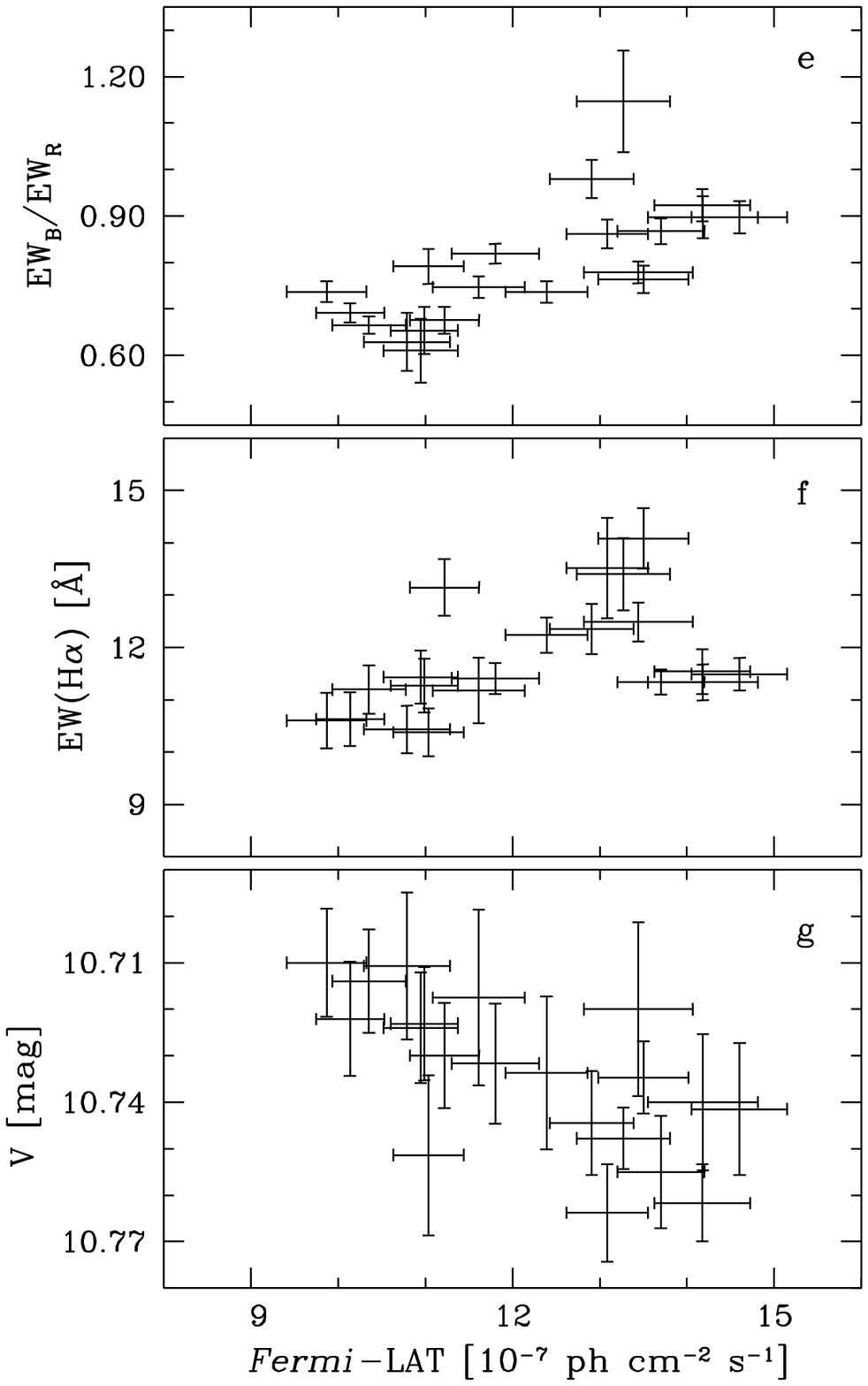}    
  \caption[]{Various parameters plotted versus $Fermi$-LAT flux.            
     Plotted  from up to down:
     2.25 GHz,
     $\Delta V_p$, 
     EW(B),
     EW(R),  
     the ratio $EW_B/EW_R$,   
     EW(\Ha),  and 
     V magnitude.  
     Each point represents 1 bin with size 0.05  (see also Fig.\ref{f1...} and Table~\ref{tab.corr}).
    }  
\label{f2...}      
\end{figure}	     

\section{Orbital variability of different parameters}

The orbital phase is calculated  using  P$_{orb}= 26.4960 \pm 0.0028$ days, a value derived 
from Bayesian analysis of the radio observations (Gregory 2002).
The zero of  phase is by convention JD$_0$=2,443,366.775, the date of the first radio detection of
the star (Gregory \& Taylor 1978).

To calculate the average orbital variability, we separated the data in 20 non-overlapping bins. 
The bins are all of the same size -- 0.05 in phase each. 
In each phase bin we calculated the average value, median value, and standard deviation of the mean.
The values were calculated separately  for every data type. 
In this way we obtained 20 points per orbital period and  plotted them
in Fig.~\ref{f1...}.  The data are repeated over two cycles for clarity. 
In Fig.~\ref{f1...} from top to bottom we plot  \\
 {\bf  a- } radio flux density at  2.25 GHz (in Jy),   \\
 {\bf  b- } radio flux density at  8.3 GHz (in Jy),  \\
 {\bf  c- } the distance between the peaks of the \Ha\ emission line, $\Delta V_p$, in \kms, \\
 {\bf  d- } equivalent width of the blue hump of \Ha,  EW(B), in \AA,\\
 {\bf  e- } equivalent width of the red hump of \Ha,  EW(R), in \AA, \\ 
 {\bf  f- } the dimensionless ratio between equivalent widths of the blue and red humps of \Ha\ $EW_B/EW_R$,  \\
 {\bf  g- } the total equivalent width of \Ha\ emission line, EW(\Ha ), in \AA, \\ 
 {\bf  h- } the optical  (Johnson) V band magnitude, \\
 {\bf  k- } the {\it Fermi}-LAT photon flux  in the 0.1-300 GeV energy range (in units  $10^{-7}$ photons~cm$^{-2}$~s$^{-1}$). \\

In this figure, it can be seen that   the minimum of  the $\gamma$-ray production 
is  around the time of the apastron and after it. This corresponds to the maximum 
of the optical V brightness. 
At the same orbital phases  a  very pronounced minimum is visible in EW(\Ha ), EW(B), and $EW_B/EW_R$.



\begin{table}
\caption{Spearman's (rho) correlation test results.}   
\centering                             
\begin{tabular}{lrlllllr}        
\hline\hline                           
 &  &  \\                         
Parameters           &  \multicolumn{2}{c}{Spearman test}     &   Result      & Fig.      \\ 	
                         &  coeff.       & p-value            &                           \\    
 \\
 \hline 
 \\
$\gamma$ -  radio       &   0.12       &  $6.3 \cdot 10^{-1}$	 &		   & 2a   \\
$\gamma$ - $\Delta V_p$ &  -0.32       &  $1.6 \cdot 10^{-1}$	 & 	           & 2b   \\
$\gamma$ - $EW(B)$      &   0.80       &  $2.1 \cdot 10^{-5}$	 &   highly sign.  & 2c   \\       
$\gamma$ - $EW(R)$      &  -0.16       &  $4.7 \cdot 10^{-1}$	 & 	           & 2d   \\ 	    
$\gamma$ - $EW_B/EW_R$  &   0.76       &  $8.8 \cdot 10^{-5}$    &   highly sign.  & 2e   \\	 
$\gamma$ - EW(\Ha)      &   0.58       &  $6.7 \cdot 10^{-3}$	 &   significant   & 2f   \\	 
$\gamma$ - $V$	        &   0.69       &  $7.0 \cdot 10^{-4}$	 &   highly sign.  & 2g   \\	 
  \\
$EW(B)$ - $V$           &   0.65       &  $2.1 \cdot 10^{-3}$	 &   significant   &      \\	 
$EW(R)$ - $V$           &  -0.23       &  $3.2 \cdot 10^{-1}$    &                 &      \\
EW(\Ha) - $V$           &   0.45       &  $4.6 \cdot 10^{-2}$	 &                 &      \\      
\\ 
\hline  
\multicolumn{3}{l}{\footnotesize{$\gamma = Fermi$-LAT flux.}}                                
\label{tab.corr}
\end{tabular}
\end{table}

\subsection{Phase lags}

\label{lag}

In Fig.~\ref{f1...} it is visible that the averaged radio fluxes peak at phase $\sim \! 0.7$, while 
$V$ magnitude peaks at phase $\sim \! 0.9$,  EW(\Ha) at  phase $\sim \! 0.5$, 
$EW(B)$   at phase  $\sim \! 0.4$, $Fermi$-LAT flux at phase  $\sim \! 0.3$. 

To estimate the phase shift between the different bands,  we used a cross correlation function (CCF) with 
the Fermi-LAT flux as a reference. 
The delay of the other parameters is as follows: radio flux at 2.25~GHz is delayed by 0.36, 
radio flux at 8.3 GHz is delayed by 0.30, 
$V$ band brightness is delayed by 0.47, 
EW(\Ha) is delayed by 0.14 , 
$EW(B)$ is delayed by 0.06, 
and $EW(R)$ is delayed by 0.28.  
The typical error of these shifts is $\pm 0.03$.

\subsection{Correlations: \Ha ,  $V$ magnitude, and radio  versus $\gamma$-rays}

The various optical and radio parameters are plotted versus $Fermi$-LAT flux in Fig.~\ref{f2...}.
In each panel there are  20 data points, plotted with their errors.  
For each panel of this figure we performed Spearman's (rho) rank correlation test.
The results of the test (correlation coefficient and p-value)  
are summarized in Table~\ref{tab.corr}, where 
the first column lists the correlated parameters,
and the second  the correlation coefficient and its significance ($p$-value). The third column
notes our result, and if no correlation is detected it is empty. The fourth
column refers to the figure where the data are plotted.

The highest correlations are between $Fermi$-LAT and $EW(B)$, as well as $Fermi$-LAT and $EW_B/EW_R$.
It is worth noting  that  $EW_B/EW_R$ is one of the \Ha\ emission parameters where 
the orbital period is most visible (Zamanov et al. 2013). The correlation is also 
highly significant ($p<0.001$) for  $\gamma$-rays and  $EW(B)$ and 
for  $\gamma$-rays  and  the optical V band magnitude. 
Between   $\gamma$-rays  and the optical V band,  there is  an anti-correlation, 
and the optical brightness decreases when the  $\gamma$ flux increases. 
There is also a correlation  between  $Fermi$-LAT flux and total $EW($\Ha$)$, which  seems to be  
significant ($p \le 0.01$). 
The statistical significance of the correlations, with chance probability values
$p \approx 10^{-4}$, indicates a relationship between the \Ha\ and $\gamma$-ray emission processes. 

It is worth noting that
{\bf (i)} there is no correlation between the EW(\Ha) and V. However, 
a correlation does exist between $EW(B)$ and V brightness. 
{\bf (ii)} If we use the phase lags to match up the orbital modulation, the Spearman test gives
a worse  result for  $\gamma - EW(B)$ (0.76, $p=1.0 \cdot 10^{-4}$), 
better result for  $\gamma$ - $EW_B/EW_R$  (0.78, $p=4.1 \cdot 10^{-5}$),
considerably better result for  $\gamma$ - EW(\Ha) (0.80 $p=2.2 \cdot 10^{-5}$), 
and an even better result for $\gamma$ - $V$ (0.75 $p=1.5 \cdot 10^{-4}$), in comparison with the values 
in Table~1.

%
%

\section{Discussion and conclusions}

The type of the secondary  in \lsi\ is still unknown. 
Several  models have been proposed
for the nature of the compact object in \lsi\ 
several  models have been proposed: 
an accreting black hole launching  relativistic jets (microquasar, e.g., Massi et al. 2012), 
rotation-powered pulsar (Dubus 2013), 
ejector-propeller (Zamanov et al. 2001), 
and accretor-ejector model (Maraschi \& Treves 1981). 
The properties of the short bursts recently observed
are typical of those shown by high magnetic field neutron stars (magnetars), so they 
provide one more indication  of  neutron star (Papitto, Torres \& Rea \ 2012). 
During the ejector stage (pulsar) the gamma ray emission is thought to originate in the shock front at the boundary of the pulsar and
stellar winds  and/or inverse Compton process. 
Electrons and hadrons can also be accelerated to relativistic energies by 
a propeller-acting neutron star (accretion onto the magnetosphere of  magnetar). These relativistic particles 
will produce $\gamma$-ray and neutrino emission (see Bednarek 2011, and references therein).


The \Ha\ emission in Be stars and Be/X-ray binaries is coming from a Keplerian  disk 
around the Be star (e.g., Hanuschik et al. 1988). This circumstellar disk also supplies the material that feeds 
the accretion  onto the X-ray pulsars in  the Be/X-ray binaries. 

{\bf Phase lags:} We detected and calculated the phase lags between $\gamma$-rays from one side 
and optical, \Ha, and radio parameters from the other side (see Sect.\ref{lag} and Fig.\ref{f1...}). 
Chernyakova et al. (2012) used simultaneous X-ray and radio observations to show that periodic radio flares 
always lag behind the X-ray flare by $\Delta\phi \simeq 0.2$,  
a behavior predicted  by the ejector - propeller model. 
The radio outbursts are probably due to an expansion  of a synchrotron emitting source (plasmon), 
with a prolonged injection of energetic particles (Paredes et al. 1991). 
In this model the phase shifts between the high-energy emission and other bands are probably 
connected with the ejection of relativistic wind (or jets) from the compact object, 
the  appearance, and the  expansion of the plasmon, 
which achieves the maximum of the radio flux  a few (2 - 8) days after its appearance.

{\bf Correlations:}
We detected  a highly significant anti-correlation between $Fermi$-LAT flux and optical V brightness.  
This could be due to changes in the ionization in the Be circumstellar disk 
in response to the high-energy emission, which changes the opacity and the emission.
The gamma-ray bright blazars in the  sample of Bonning et al. (2012) 
have optical emission correlated with gamma-rays. Bonning et al. (2012) suggest that this strongly supports
leptonic models for the gamma-ray production. 
We have the opposite situation, which is hard to reconcile with leptonic models based on the inverse Compton origin 
of the $Fermi$-LAT photons from \lsi.


The $Fermi$-LAT flux achieves the maximum at about the time of the periastron passage.
The highest values of the $EW(B)$ are reached in the phase interval 0.3 - 0.6, 
The highest values of $EW_B/EW_R$ are about the orbital phase 0.40. 
The highest values of the total EW(\Ha) are also reached in the phase interval 0.3 - 0.6. 
That  the $Fermi$-LAT flux correlates with $EW(B)$ and  $EW_B/EW_R$,
and not with EW(R) indicates that  the high-energy emission  does not
influence  all the Be disk but only the  vicinity around the compact object. 


For rotationally dominated profiles the peak separation can be regarded as a measure of 
the outer radius (e.g., Hanuschik et al. 1988).
Because there is no correlation of  $Fermi$/LAT flux  and $\Delta V_p$, it means that the  
the size of the  \Ha\ emitting disk does not respond to the changes of the $\gamma$-ray flux.
This in agrees with the above that only the surroundings of the compact object star are involved.

{\bf Conclusions:} 
We detected  highly significant correlations (chance probability value $\approx 10^{-4}$ ) 
between the orbital modulation of the blue hump of \Ha\ emission line and $Fermi$-LAT flux
and between the ratio $EW_B/EW_R$  and $Fermi$-LAT flux, 
as well as an anti-correlation between V band brightness and $Fermi$-LAT  flux.
This implies a direct link between the \Ha\ and $\gamma$-ray emission processes.

\begin{acknowledgements} We are very grateful to the referee, E. Grundstrom, for valuable comments.
 This work was  partially  supported the OP "HRD", ESF, and Bulgarian  Ministry of Education and Science
 (BG051PO001-3.3.06-0047). 
 JM  acknowledges support by grant AYA2010-21782-C03-03 from the Spanish Government, and
 Consejer\'{\i}a de Econom\'{\i}a, Innovaci\'on y Ciencia of Junta de Andaluc\'{\i}a as research group FQM-322, 
 as well as FEDER funds.
\end{acknowledgements}


\begin{table*}
\caption{The averaged  orbital variability (20 bins)  of \lsi.}   
\centering                             
\begin{tabular}{cccccccc}        
\hline 
             &   \\                         
phase bin    &   $EW(B)$        & $EW(R)$  	 &  $EW_B/EW_R$     &     EW(\Ha)      &      $V$	  &     $\Delta V_p$  &  \\
& \\
             &   [\AA]          & [\AA]          &  [\AA]           &  [\AA]           &   [mag]          &  [\kms ]         &   \\
& \\
0.00-0.05    &   4.780\e 0.306  & 6.392\e 0.343  &  0.7468\e 0.0229 &  11.173\e 0.629  &  10.7175\e0.0189 &  336.07\e  4.28  &   \\
0.05-0.10    &   5.138\e 0.193  & 6.256\e 0.135  &  0.8190\e 0.0212 &  11.409\e 0.295  &  10.7317\e0.0129 &  327.17\e  2.43  &   \\
0.10-0.15    &   4.558\e 0.304  & 5.775\e 0.256  &  0.7917\e 0.0377 &  10.380\e 0.460  &  10.7514\e0.0172 &  338.61\e  3.22  &   \\
0.15-0.20    &   5.241\e 0.140  & 6.096\e 0.151  &  0.8672\e 0.0271 &  11.337\e 0.237  &  10.7550\e0.0121 &  341.30\e  3.24  &   \\
0.20-0.25    &   5.415\e 0.201  & 6.074\e 0.162  &  0.8974\e 0.0351 &  11.489\e 0.311  &  10.7415\e0.0142 &  344.92\e  4.01  &   \\
0.25-0.30    &   5.342\e 0.270  & 5.990\e 0.161  &  0.8969\e 0.0450 &  11.332\e 0.343  &  10.7400\e0.0146 &  339.69\e  4.96  &   \\
0.30-0.35    &   5.503\e 0.219  & 6.036\e 0.252  &  0.9232\e 0.0349 &  11.539\e 0.428  &  10.7617\e0.0083 &  328.42\e  4.34  &   \\
0.35-0.40    &   7.245\e 0.497  & 6.517\e 0.398  &  1.1465\e 0.1090 &  13.402\e 0.690  &  10.7478\e0.0066 &  312.95\e  7.86  &   \\
0.40-0.45    &   6.163\e 0.313  & 6.326\e 0.227  &  0.9797\e 0.0408 &  12.354\e 0.480  &  10.7445\e0.0112 &  321.89\e  5.93  &   \\
0.45-0.50    &   6.247\e 0.473  & 7.274\e 0.513  &  0.8613\e 0.0304 &  13.521\e 0.962  &  10.7638\e0.0105 &  319.35\e  11.4  &   \\
0.50-0.55    &   6.274\e 0.356  & 8.173\e 0.274  &  0.7638\e 0.0297 &  14.084\e 0.580  &  10.7347\e0.0078 &  324.87\e  10.0  &   \\
0.55-0.60    &   5.565\e 0.206  & 7.194\e 0.237  &  0.7787\e 0.0234 &  12.490\e 0.371  &  10.7200\e0.0187 &  336.06\e  7.74  &   \\
0.60-0.65    &   5.231\e 0.201  & 7.147\e 0.173  &  0.7366\e 0.0236 &  12.238\e 0.337  &  10.7337\e0.0165 &  336.23\e  4.30  &   \\
0.65-0.70    &   5.212\e 0.218  & 7.935\e 0.384  &  0.6758\e 0.0293 &  13.148\e 0.545  &  10.7300\e0.0113 &  353.55\e  4.16  &   \\
0.70-0.75    &   4.563\e 0.273  & 6.637\e 0.363  &  0.6536\e 0.0507 &  11.270\e 0.517  &  10.7231\e0.0122 &  345.93\e  4.87  &   \\
0.75-0.80    &   4.829\e 0.310  & 6.892\e 0.226  &  0.6101\e 0.0685 &  11.433\e 0.507  &  10.7240\e0.0119 &  339.71\e  5.57  &   \\
0.80-0.85    &   4.416\e 0.265  & 6.274\e 0.251  &  0.6291\e 0.0622 &  10.435\e 0.452  &  10.7107\e0.0158 &  344.06\e  5.48  &   \\
0.85-0.90    &   4.485\e 0.226  & 6.740\e 0.289  &  0.6651\e 0.0184 &  11.197\e 0.464  &  10.7140\e0.0111 &  333.27\e  4.36  &   \\
0.90-0.95    &   4.480\e 0.232  & 6.478\e 0.273  &  0.6915\e 0.0203 &  10.626\e 0.516  &  10.7221\e0.0123 &  343.30\e  6.78  &   \\
0.95-1.00    &   4.628\e 0.274  & 6.212\e 0.245  &  0.7373\e 0.0221 &  10.601\e 0.534  &  10.7100\e0.0116 &  338.96\e  4.48  &   \\
&  \\
\hline                  	       
\label{tab.apend.2} 
\end{tabular}
\end{table*}


\end{document}